\documentclass[12pt]{article}

\begin{document}

\title{Conservation laws for the nonlinear Schr\"{o}dinger equation in
Miwa variables.}

\author{G.M. Pritula and V.E. Vekslerchik
\thanks{Regular Associate of the
   Abdus Salam International Centre for Theoretical Physics,
   Trieste, Italy}
\\
\\
\normalsize\it
  Institute for Radiophysics and Electronics,  \\
\normalsize\it
  National Academy of Sciences of Ukraine,     \\
\normalsize\it
  Proscura Street 12, Kharkov 310085, Ukraine.}

\date{\today}
\maketitle

\begin{abstract}
A compact expression for the generating function of the constants of
motion for the nonlinear Schr\"{o}dinger equation is derived using the
functional representation of the AKNS hierarchy.
\end{abstract}

\section{Introduction.}

In the present note we would like to discuss once more the conservation
laws for the nonlinear Schr\"{o}dinger equation (NLSE),

\begin{eqnarray}
i{\partial q \over \partial t} +
{\partial^{2} q \over \partial x^{2}} +
2 q^{2} r  &=& 0
\\
-i{\partial r \over \partial t} +
{\partial^{2} r \over \partial x^{2}} +
2 q r^{2} &=& 0
\end{eqnarray}

Existence of an infinite number of conserved quantities is an
characteristic feature of integrable partial differential equations (PDE),
and this question in context of the NLSE was discussed in the very first
works devoted to this model, where authors established integrability of
the NLSE and elaborated the corresponding inverse scattering transform
(IST). The IST is based on a representation of the equation in question as
a compatibility condition for a overdetermined linear system (the
so-called zero curvature representation) which in the case of the NLSE can
be written as

\begin{eqnarray}
{\partial \over \partial x} \Psi(x,t;\lambda) &=&
  U(x,t;\lambda) \Psi(x,t;\lambda)
\label{sp_nlse}
\\
{\partial \over \partial t} \Psi(x,t;\lambda) &=&
V(x,t;\lambda) \Psi(x,t;\lambda)
\label{evol_nlse}
\end{eqnarray}
Here $\Psi$ is a $2$-column, the matrix $U$ is given by

\begin{equation}
U = i \pmatrix{ \lambda & r \cr q & -\lambda }
\label{U}
\end{equation}
and $V$ is some $2 \times 2$ matrix which is a second order polynomial in
$\lambda$ (in what follows we will not need its explicit form).

The $U$-$V$ representation (\ref{sp_nlse}), (\ref{evol_nlse}) of the NLSE is enough
to obtain answers to a wide range of questions related to this equation.
In particular, one can derive from (\ref{sp_nlse}), (\ref{evol_nlse}) an
infinite number of integrals of motion of the NLSE. This can be done as
follows (here we will only outline some key moments, more elaborated
description of the IST one can find in various textbooks on this topic,
as, e.g., \cite{AS,N,FT}).  Introducing the so-called Jost functions
$\Psi_{\pm}$ of the scattering problem (\ref{sp_nlse}) (i.e. solutions of
(\ref{sp_nlse}) satisfying different boundary conditions) and the
scattering matrix $T(x,t;\lambda)$, $\Psi_{+} = \Psi_{-} T$, one can
obtain from the zero-curvature representation that the diagonal elements
of the $2 \times 2$ matrix $T$ do not depend on time. Hence they
(or their logarithms) can be used as generating functions of constants of
motion (the later are the coefficients of the power series in $\lambda$
($\lambda^{-1}$) of the former).

Another approach stems from the viewpoint when one considers an
integrable equation (the NLSE in our case) as a member of an integrable
hierarchy (the AKNS in our case).  The $U$-$V$ representation of the
equations of the AKNS hierarchy can be written as

\begin{eqnarray}
{\partial \over \partial x} \Psi &=& U \Psi
\label{sp_akns}
\\
{\partial \over \partial t_{n}} \Psi &=& V_{n} \Psi
\label{evol_akns}
\end{eqnarray}
where $U$ is given again by (\ref{U}) and $V_{n}$'s are different matrices
($V_{n}$ is some $n$th order polynomial in $\lambda$). All equations of
the hierarchy are compatible and one can solve them simultaneously,
i.e.  one can think of $q$ and $r$ as functions of an infinite set of
variables $q,r=q,r(t_{1},t_{2},t_{3},...)$ with the evolution with respect
to $t_{k}$ being given by the $k$th NLSE ($k$th member of the AKNS
hierarchy). In some situations such standpoint leads to more transparent
results, and the main aim of this note is to apply it to the question of
the description of the constants of motion of the NLSE (read constants of
motion of the AKNS hierarchy).

\section{Functional representation of the AKNS hierarchy.}

Our starting point is the so-called functional representation of the AKNS
hierarchy which can be written as

\begin{eqnarray}
i\zeta \partial_{1} q &=& q - q^{-} + \zeta^{2} q^{2} r^{-}
\label{fr_q}
\\
- i\zeta \partial_{1} r &=& r - r^{+} + \zeta^{2} q^{+} r^{2}
\label{fr_r}
\end{eqnarray}
Here $q$ and $r$ are functions of an infinite set of times,
$q = q(\mathrm{t})$, $r = r(\mathrm{t})$,

\begin{equation}
f(\mathrm{t}) = f(t_{1},t_{2},t_{3},...)
\end{equation}
$\partial_{1} = \partial / \partial t_{1}$ and the designation $f^{\pm}$
stands for the function with shifted arguments (Miwa shifts),

\begin{eqnarray}
f^{\pm}
&=&
f(\mathrm{t} \pm i[\zeta])
\\
&=&
f(t_{1} \pm i\zeta, t_{2} \pm i\zeta^{2}/2, t_{3} \pm i\zeta^{3}/3,...)
\end{eqnarray}

This equation, which can be termed as 'AKNS hierarchy in Miwa variables',
may be derived in different ways. First this can be done by careful
analysis of the linear problems (\ref{sp_akns}), (\ref{evol_akns}). One
can find the shifts $t_{k} \to t_{k} \pm i\zeta^{k}/k$ in some textbooks
on the IST (see, e.g., chapters 3,4 of the book \cite{N}). Another way is
to use the (generalized) Hirota's bilinear identities which are one of the
most important formulae of the Kyoto school approach to the integrable
systems. Explicitly equations (\ref{fr_q}) and (\ref{fr_r}) have been
written down in the paper \cite{V99}. One should also mention paper
\cite{BK} where similar representation has been derived for the
Davey-Stewartson system, from which one can easily obtain one for the AKNS
hierarchy. We will not repeat here derivation of (\ref{fr_q}) and
(\ref{fr_r}) and only demonstrate that the first equations of the AKNS
hierarchy (the NLSE in particular) can be easily obtained from them.
Indeed, using the multidimensional Taylor series for $f^{\pm} =
f(\mathrm{t} \pm i[\zeta])$,

\begin{equation}
f^{\pm} =
f \pm i\zeta\partial_{1}f
+
{\zeta^{2} \over 2}
\left( \pm\, i\partial_{2}f - \partial_{11}f \right)
+
{\zeta^{3} \over 6}
\left(
  \pm\, 2i\partial_{3}f - 3\partial_{21}f \mp\, i\partial_{111}f \right)
+
...
\end{equation}
(here $\partial_{k}$ stands for $\partial / \partial t_{k}$,
$\partial_{jk}$ for $\partial^{2} / \partial t_{j} \partial t_{k}$, etc)
and expanding equations (\ref{fr_q}), (\ref{fr_r}) in power series in
$\zeta$ one will obtain that the functions $q$ and $r$ satisfy an infinite
number of PDEs. The first non-trivial equations (the $\zeta^{2}$ terms),

\begin{eqnarray}
i\partial_{2} q + \partial_{11} q + 2 q^{2} r  &=& 0
\label{fr_1_q}
\\
-i\partial_{2} r + \partial_{11} r + 2 q r^{2} &=& 0
\label{fr_1_r}
\end{eqnarray}
are nothing else than the NLSE. Hereafter we will identify variables $x$,
$t$ and $t_{1}$, $t_{2}$

\begin{equation}
x = t_{1},
\qquad
t = t_{2}
\end{equation}
The next equations

\begin{eqnarray}
2\partial_{3} q - 3i\partial_{21} q
- \partial_{111} q - 6 q^{2} \partial_{1}r &=& 0
\\
2\partial_{3} r + 3i\partial_{21} r
- \partial_{111} r - 6r^{2} \partial_{1}q &=& 0
\end{eqnarray}
can be rewritten using (\ref{fr_1_q}), (\ref{fr_1_r}) as

\begin{eqnarray}
\partial_{3} q + \partial_{111} q + 6 qr \partial_{1}q &=& 0
\label{fr_2_q}
\\
\partial_{3} r + \partial_{111} r + 6 qr \partial_{1}r &=& 0
\label{fr_2_r}
\end{eqnarray}
These are the third-order NLSE. Thus one can view (\ref{fr_q}),
(\ref{fr_r}) as a 'condensed' form of the AKNS hierarchy.

The key moment is that, if we deal not only with solutions of the NLSE,
$q,r(x,t)$ but consider them as solutions of all equations of the
hierarchy, $q=q(t_{1},t_{2},...)$, $r=r(t_{1},t_{2},...)$, then we can
formally solve the auxiliary linear problem. Indeed, it follows from
(\ref{fr_q}), (\ref{fr_r}) that the $2 \times 2$ matrix

\begin{equation}
\Psi =
\pmatrix{ 1 & -\zeta r^{-} \cr \zeta q^{+} & 1 }
\pmatrix{ \exp(iu_{1}) & 0 \cr 0 & \exp(-iu_{2}) }
\label{BAF}
\end{equation}
where

\begin{eqnarray}
u_{1} &=& { x \over 2\zeta } + \zeta \int \mathrm{d} x \; q^{+} r
\\
u_{2} &=& { x \over 2\zeta } + \zeta \int \mathrm{d} x \; q r^{-}
\end{eqnarray}
solves (\ref{sp_akns}) with $\lambda = (2\zeta)^{-1}$. Hence we can now 
rewrite the results which were presented in terms of solutions of 
(\ref{sp_akns}) (i.e. in terms of the Jost or Baker-Akhiezer functions) in 
terms of $q$, $r$ themselves. This is also valid for the generating 
function of the integrals of motion.

Of course, matrix (\ref{BAF}) is a {\it formal} solution of (\ref{sp_akns})
and one must be ready to face some problems when, e.g., one will try to
construct the Jost functions (i.e. to satisfy some boundary conditions).
However, for our purposes this is not an obstacle. Moreover, we will not
repeat the 'classical' algorithm, Jost functions $\Psi_{\pm}(\lambda)$
$\to$  scattering matrix $T(\lambda)$ $\to$ generating function
$\ln T_{11}(\lambda)$. Knowing the answer, we will first present the final
result, and then, using the functional representation (\ref{fr_q}),
(\ref{fr_r}) of the AKNS hierarchy, will prove it.

\section{Conservation laws.}

The main result of this work can be presented as follows: the function

\begin{equation}
J(\mathrm{t},\zeta) = q(\mathrm{t} + i[\zeta]) r(\mathrm{t})
\label{main}
\end{equation}
is the generating function for the constants of motion.

Indeed, it follows from (\ref{fr_1_q}), (\ref{fr_1_r})  that

\begin{equation}
i {\partial \over \partial t}  q^{+}r =
{\partial \over \partial x}
  \left(
  q^{+}{\partial r \over \partial x} -
  {\partial q^{+} \over \partial x} r
  \right)
- 2 q^{+}r \left( q^{+}r^{+} - qr \right)
\end{equation}
Using again (\ref{fr_1_q}), (\ref{fr_1_r}), this time with shifted
arguments, one can easily get

\begin{equation}
i\zeta {\partial \over \partial x} q^{+} r =  q^{+}r^{+} - qr
\end{equation}
which leads to

\begin{equation}
{\partial \over \partial t} J(\zeta) =
{\partial \over \partial x} F(\zeta)
\end{equation}
where

\begin{equation}
F = i
{\partial q^{+} \over \partial x} r -
q^{+} {\partial r \over \partial x} -
\zeta \left( q^{+}r \right)^{2}
\end{equation}
(recall that $x = t_{1}$ and $t = t_{2}$).

Thus we have obtained an infinite number of divergent-like conservation
laws

\begin{equation}
{\partial \over \partial t} J_{m} =
{\partial \over \partial x} F_{m}
\label{div_n}
\end{equation}
where $J_{m}$'s and $F_{m}$'s are coefficients of the Taylor series for
$J(\zeta)$ and $F(\zeta)$

\begin{eqnarray}
J(\zeta) &=&  \sum_{m=0}^{\infty} J_{m} \zeta^{m}
\\
F(\zeta) &=&  \sum_{m=0}^{\infty} F_{m} \zeta^{m}
\end{eqnarray}
Some first of the conserved densities are given by

\begin{eqnarray}
J_{0}  &=& qr
\\
J_{1}  &=& i {\partial q \over \partial x} \, r
\\
J_{2}  &=&
  {1 \over 2} \left( i \partial_{2} q - \partial_{11} q \right) r
=
  - \left( {\partial^{2} q \over \partial x^{2} } + q^{2}r \right) r
\\
J_{3}  &=&
\left( 
  {i \over 3} \, \partial_{3} q - 
  {1 \over 2} \, \partial_{21} q - 
  {i \over 6} \, \partial_{111} q 
\right) r
=
-i \left( 
  {\partial^{3} q \over \partial x^{3} } + 
  4qr {\partial q \over \partial x } + 
  q^{2} {\partial r \over \partial x} 
\right) r
\end{eqnarray}
Note that to present $J_{m}$ for $m=2,...$ in a standard way, i.e. in
terms of $q$, $r$ and their derivatives with respect to $x$, one has to
use evolution equations of the hierarchy (\ref{fr_2_q}), (\ref{fr_2_r})
and higher. However, it is possible not to use these equations but instead
to 'iterate' the identity (\ref{fr_q}), which can be rewritten as

\begin{eqnarray}
q^{+}
&=& q + i\zeta q^{+}_{x} - \zeta^{2} \left( q^{+} \right)^{2} r
\label{fr_q_alt}
\\
&=& q + i\zeta q_{x}
- \zeta^{2} \left[ q^{+}_{xx} + \left( q^{+} \right)^{2} r \right]
- i\zeta^{3} \left[ \left( q^{+} \right)^{2} r \right]_{x}
\\
&=& \dots
\end{eqnarray}
or to return to the traditional inverse scattering scheme: it follows from
(\ref{fr_q_alt}) that $J(\mathrm{t},\zeta)$ satisfies

\begin{equation}
J = qr + i\zeta r {\partial \over \partial x} {J \over r} - \zeta^{2}J^{2}
\label{Riccati}
\end{equation}
which leads to the recurrence relation

\begin{eqnarray}
&&
J_{0} = qr,
\quad
J_{1} = q_{x}r
\\
&&
J_{m+1} =
r {\partial \over \partial x} {J_{m} \over r} -
\sum_{l=0}^{m-1} J_{l} J_{m-1-l},
\qquad
m \ge 1
\end{eqnarray}
One can easily identify these equations with the standard for the inverse
scattering approach equations for the generating function.
Thus, the main result (\ref{main}) of this paper can be interpreted as
follows. Equation (\ref{Riccati}), if considered as an ordinal
differential equation, is the famous Riccati equation which cannot be
solved explicitly for arbitrary functions $q$ and $r$. However in our case
$q$ and $r$ are not arbitrary but related by an infinite number of PDEs of
the hierarchy, and it turns out that in this situation, though these
restrictions do not determine $q$ and $r$ uniquely, equation
(\ref{Riccati}) can be solved formally and this solution is given by
(\ref{main}).

Equations (\ref{div_n}) can be rewritten as

\begin{equation}
{\partial \over \partial t} I_{m} =  0
\end{equation}
where $I_{m}$'s are integrals of the densities $J_{m}$. In the case when
$q$ and $r$ vanish (sufficiently rapidly) as $x \to \pm\infty$ $I_{m}$'s
are given by

\begin{equation}
I_{m} =  \int_{-\infty}^{\infty} \mathrm{d}x \; J_{m}
\end{equation}
In the periodical case, $q,r(x+L) = q,r(x)$,

\begin{equation}
I_{m} =  \int_{0}^{L} \mathrm{d}x \; J_{m}
\end{equation}
while in the case of non-trivial boundary conditions, say the
finite-density ones, the integral in the right-hand side should be in some
way regularized.

\section{Conclusion.}

To conclude we would like to note the following.
The aim of this note is twofold. First we want to demonstrate that the
Kyoto school approach can be used not only to reveal and study some
mathematical structures behind integrable equations, but also to solve
some 'practical' problems, as one discussed above. Another purpose of this
paper is to attract attention to the functional representation of the AKNS
hierarchy. This system is one of the first studied integrable models for
which the IST has been developed, and since 70's till now the IST is the
main tool to study the NLSE and related problems. As to the methods which
were developed later for such equations as, e.g., KP and 2D Toda
equations, for our knowledge, their application to the AKNS hierarchy is
rather limited. At the same time the formulation of a problem in terms of
the functional equations seems be perspective for, e.g., developing some
perturbation schemes for various NLSE-related problems.



\begin{thebibliography}{**}
\bibitem{AS}
  Ablowitz M.J. and Segur H.
  {\it Solitons and the Inverse Scattering Transform.}
  1981 (Philadelphia: SIAM).

\bibitem{N}
  Newell A.C.
  {\it Solitons in Mathematics and Physics.}
  1985 (Philadelphia: SIAM).

\bibitem{FT}
  Faddeev L.D. and Takhtajan L.A.
  {\it Hamiltonian Methods in the Theory of Solitons.}
  1987 (Berlin: Springer).

\bibitem{V99}
  Vekslerchik V.E.
  Functional representation of the Ablowitz-Ladik hierarchy. II.
  {\tt solv-int/9812020}

\bibitem{BK}
  Bogdanov L.V., Konopelchenko B.G.
  Analytic-bilinear approach to integrable hierarchies. II. Multicomponent
  KP and 2D Toda lattice hierarchies.
  {\it J.Math.Phys.}, 1998, {\bf 39}, 4701-4728.

\end{thebibliography}
\end{document}